\documentclass[11pt,letterpaper]{article}
\pdfoutput=1
\usepackage{jheppub}
\usepackage{bbm}
\usepackage{mathrsfs}
\usepackage{slashed}
\usepackage{caption}
\usepackage{epstopdf}
\usepackage[normalem]{ulem}
\usepackage[bottom]{footmisc}
\usepackage{subcaption}
\usepackage{bbold}
\usepackage{titlesec}
\usepackage{threeparttable}
\usepackage{booktabs}
\usepackage{changepage}
\usepackage[utf8]{inputenc}

\usepackage{grffile}

\usepackage{graphicx}  
\usepackage{dcolumn}   
\usepackage{bm}        
\usepackage{amssymb}   
\usepackage{setspace}
\usepackage{amsmath, amssymb, setspace}
\usepackage{array}
\usepackage{booktabs}
\usepackage{caption}
\usepackage{indentfirst}
\usepackage{float}
\usepackage{lmodern}
\usepackage{multirow}
\usepackage{soul}
\usepackage[normalem]{ulem}

\usepackage{braket}
\usepackage{comment}

\usepackage[draft]{pgf}

\usepackage{adjustbox} 
\newcommand{\DoBox}[1]{\begin{center}
\color{red}\fbox{
\begin{minipage}{0.9\textwidth}

\end{minipage}}
\end{center}}

\newlength{\myimageoversize}
\newsavebox{\myimage}

\titleformat{\subsubsection}
 {\normalfont\fontsize{12}{17}\itshape}{\thesubsubsection}{1em}{}

\title{\huge{Note on Thermalization of Non-resonantly Produced Sterile Neutrinos}}

\author[a]{Graciela B. Gelmini,}
\author[a]{Philip Lu}
\author[a]{and Volodymyr Takhistov}

\affiliation[a]{Department of Physics and Astronomy, University of California, Los Angeles\\
 Los Angeles, CA 90095-1547, USA}

\emailAdd{gelmini@physics.ucla.edu}
\emailAdd{philiplu11@gmail.com}
\emailAdd{vtakhist@physics.ucla.edu}

\begin{document}

\abstract{
Using an analytic treatment, we discuss the parameter regions of large active-sterile neutrino mixing angles where sterile neutrinos produced in non-resonant flavour oscillations can approach thermalization in several cosmologies. We show that thermalization affects only large active-sterile neutrino mixing already rejected by different limits. Hence, the allowed sterile neutrino parameter regions are unaffected.}

\maketitle
 
\section{Introduction}

In Ref.~\cite{Gelmini:2019wfp} (from here on Paper I), as well as in its prior abbreviated companion paper~\cite{Gelmini:2019esj}, we have considered the cosmological dependence of non-resonantly produced sterile neutrinos. We discussed there the sensitivity of sterile neutrino production to cosmologies that differ from the standard radiation dominated cosmology (STD) before Big Bang Nucleosynthesis (BBN), specifically before the temperature of the Universe was 5 MeV. The lower limit on the highest temperature of the radiation-dominated epoch in which BBN happened is close to 5 Mev~\cite{Hasegawa:2020ctq, Hasegawa:2019jsa, deSalas:2015glj, DeBernardis:2008zz, Hannestad:2004px, Kawasaki:2000en, Kawasaki:1999na}. Thus, the cosmological evolution in the Universe before the temperature of the Universe was about 5 MeV is unknown and could differ from the STD. Alternative cosmologies can often appear in motivated theories. As examples, we have considered two distinct Scalar-Tensor models (ST1~\cite{Catena:2004ba} and ST2~\cite{Catena:2007ix}), Kination (K)~\cite{Spokoiny:1993kt,Joyce:1996cp,Salati:2002md,Profumo:2003hq,Pallis:2005hm} as well as Low Reheating Temperature (LRT) scenario~\cite{Gelmini:2004ah} (see also 
e.g.~\cite{Gelmini:2006pw, Gelmini:2006pq, Gelmini:2008fq, Yaguna:2007wi, deSalas:2015glj, Abazajian:2017tcc, Hasegawa:2019jsa, Hasegawa:2020ctq}), besides the STD cosmology (see Paper I for a detailed description of the different models). We discussed how the resulting limits and regions of interest in the mass-mixing ($m_s, \sin^2 2\theta$) plane are affected for a sterile neutrino of mass $m_s$ that is assumed to have a mixing  $\sin\theta$ only with the active electron neutrino.

In Paper I we presented a simplified treatment of sterile neutrino production for the parameter region where mixing angles are very large. There, the momentum $\vec{p}$ distribution $f_{\nu_s}(E,T)$ of sterile neutrinos of energy $E=| \vec{p} |$, which are relativistic at the production temperature $T$, is not much smaller than the distribution of active neutrinos $f_{\nu_\alpha}(E,T)$. In part of this region sterile neutrinos can thermalize in the Early Universe, so that $f_{\nu_s}(E,T)= f_{\nu_\alpha}(E,T)$. When thermalized, sterile neutrinos have the same number density of one active neutrino species, i.e. during BBN and later $\Delta N_{\rm eff} = 1$ (or close to 1, depending on entropy dilution), a value that is forbidden by present cosmological limits. While neutrino thermalization has been extensively studied with numerical methods~(e.g.~\cite{Hasegawa:2019jsa,Hannestad:2012ky}), here we analyze these effects analytically.

We show, always using analytic expressions as in our previous studies, that the regions allowed by all sterile neutrino bounds are not affected by the present considerations.

\section{Approaching Thermalization}
\label{sec:approach}
In our analysis of Paper I, we assumed $f_{\nu_s} \ll f_{\nu_\alpha}$ and thus neglected the second term on the right hand side of the Boltzmann equation
\begin{equation}
\label{eq:Boltzmann}
    \left(\frac{\partial f_{\nu_s}(E,T)}{\partial T}\right)_{\epsilon=E/T} = - \dfrac{\Gamma_s (E,T)}{ HT} [f_{\nu_\alpha}(E,T)-f_{\nu_s}(E,T)]~.
\end{equation}
where $\epsilon = E/T= |\vec{p}|/T$ is the $T$-scaled dimensionless momentum and the derivative on the left hand side is computed at constant $\epsilon$. Here $H$ is the expansion rate of the Universe, which for $T> T_{\rm tr}$ we parameterize as $H=\eta (T/T_\textrm{tr})^\beta H_\textrm{STD}$, where $H_\textrm{STD} = T^2/M_\textrm{Pl}\sqrt{8\pi^3 g_\ast(T)/90}$ is the expansion for the STD cosmology. For the non-standard cosmologies we consider, the scale factor $\eta$ and the exponent $\beta$ values are $\eta = 1$ and $\beta = 1$ for K,  $\eta = 7.4\times10^5$ and $\beta = -0.8$ for ST1, $\eta = 3.2\times10^{-2}$ and $\beta = 0$ for ST2, and they transition to the STD at $T_\textrm{tr}=$ 5 MeV (see Fig.~1 of Paper I). We also consider a LRT model with 
reheating temperature $T_\textrm{RH} = T_\textrm{tr}$.  $\Gamma_s(E,T)$ in Eq.~\eqref{eq:Boltzmann} is the conversion rate of active to sterile neutrinos,
\begin{equation} 
\label{eq:interaction}
    \Gamma ~=~ \dfrac{1}{2}  \langle P_m (\nu_\alpha \rightarrow \nu_s) \rangle \Gamma_{\alpha}   ~\simeq~ \frac{1}{4}\sin^2(2\theta_m)d_\alpha G_F^2\epsilon T^5~.
\end{equation}
where $d_\alpha = 1.27$ for $\nu_e$ and $G_F$ is the Fermi constant. In the absence of a large lepton asymmetry, the matter mixing angle is 
\begin{equation}
\label{eq:mattermixing}
    \sin^2(2\theta_m) = \frac{\sin^2(2\theta)}{\sin^2(2\theta) + \Big[\cos(2\theta)-2\epsilon T V_T/m_s^2\Big]^2}~,
\end{equation}
where for $\nu_e$ the thermal potential is $V_T = -10.88\times10^{-9}\textrm{ GeV}^{-4}$.
 
  If  $f_{\nu_s}\ll f_{\nu_\alpha}$,  $f_{\nu_s}$ can be neglected on the right hand side of Eq.~\eqref{eq:Boltzmann}, which amounts to neglecting the inverse oscillation process $\nu_s \rightarrow \nu_\alpha$. This is a good approximation for most of the large parameter space we studied with our analytic methods, 0.01 eV $< m_s < 1$ MeV and $10^{-13} < \sin^2 2 \theta < 1$. However, this approximation fails for very large mixing angles, for which  sterile neutrinos thermalize, thus  $f_{\nu_s}= f_{\nu_\alpha}$ and the right hand side of Eq.~\eqref{eq:Boltzmann} vanishes. 
 
 The ``linear" equation 
Eq.~\eqref{eq:Boltzmann} without $f_{\nu_s}$ in the right hand side, can be analytically solved for $f_{\nu_s}$, to obtain what we call now $f_{\nu_s-{\rm lin}}$ (see Paper I for a detailed discussion) for all the cosmologies we consider,
\begin{align}
\label{eq:flin}
    &f_{\nu_s-{\rm lin}}^{\rm STD}(\epsilon) = 1.04\times10^{-5}\left(\frac{\sin^2(2\theta)}{10^{-10}}\right)
    \left(\frac{m_s}{\textrm{keV}}\right) \left(\frac{g_{\ast}}{30}\right)^{-\frac{1}{2}} \left(\frac{d_\alpha}{1.27}\right) f_{\nu_\alpha}(\epsilon)~,\notag\\
    &f_{\nu_s-{\rm lin}}^{\rm K}(\epsilon) = 4.2\times10^{-7} \epsilon^{\frac{1}{3}}\left(\frac{\sin^2(2\theta)}{10^{-10}}\right)\left(\frac{m_s}{\textrm{keV}}\right)^{\frac{2}{3}} \left(\frac{g_{\ast}}{30}\right)^{-\frac{1}{2}} \left(\frac{d_\alpha}{1.27}\right) f_{\nu_\alpha}(\epsilon)~, \notag\\
    &f_{\nu_s-{\rm lin}}^{\rm ST1}(\epsilon) = 2.2\times10^{-10} \epsilon^{-0.27} \left(\frac{\sin^2(2\theta)}{10^{-10}}\right) \left(\frac{m_s}{\textrm{keV}}\right)^{1.27} \left(\frac{g_{\ast}}{30}\right)^{-\frac{1}{2}} \left(\frac{d_\alpha}{1.27}\right) f_{\nu_\alpha}(\epsilon)~,\\
    &f_{\nu_s-{\rm lin}}^{\rm ST2}(\epsilon) = 3.25\times10^{-4}\left(\frac{\sin^2(2\theta)}{10^{-10}}\right)\left(\frac{m_s}{\textrm{keV}}\right)\left(\frac{g_{\ast}}{30}\right)^{-\frac{1}{2}} \left(\frac{d_\alpha}{1.27}\right)  f_{\nu_\alpha}(\epsilon) \notag\\
    &f_{\nu_s-{\rm lin}}^{\textrm{LRT}} = 3.6\times10^{-10} \,  \epsilon \, \left(\frac{\sin^2(2\theta)}{10^{-10}}\right) \left(\frac{d_\alpha}{1.13}\right)f_{\nu_\alpha}(\epsilon)~. \notag
\end{align}

Integrating these distributions over momentum yields the corresponding ``linear" number densities $n_{\nu_s-{\rm lin}}$. Requiring the present sterile neutrino energy density not to exceed the present dark matter (DM) density $m_s n_{\nu_s-{\rm lin}}/ \rho_c < \rho_{DM}/\rho_c=\Omega_{DM}$,  where $\rho_c$ is the critical density, yields the ``old" mixing angle limits found in Paper I (in the linear approximation),
\begin{align}
\label{eq:thetalin}
        \sin^2(2\theta)_\textrm{old}^{\rm STD} &~=~ 3.58\times10^{-7}\left(\frac{m_s}{\textrm{keV}}\right)^{-2}  \left(\frac{g_{\ast}}{30}\right)^{\frac{3}{2}}\left(\frac{d_\alpha}{1.27}\right)^{-1}\left(\frac{\Omega_\textrm{DM}h^2}{ 0.12}\right)^{-1}~, \notag\\
        \sin^2(2\theta)_\textrm{old}^{\rm K} &~=~ 6.26\times10^{-6} \left(\frac{m_s}{\textrm{keV}}\right)^{-\frac{5}{3}}\left(\frac{g_{\ast}}{30}\right)^{\frac{3}{2}} \left(\frac{d_\alpha}{1.27}\right)^{-1} \left(\frac{\Omega_\textrm{DM}h^2}{ 0.12}\right)^{-1}~, \notag\\
       \sin^2(2\theta)_\textrm{old}^{\rm ST1} &~=~ 2.19\times10^{-2}\left(\frac{m_s}{\textrm{keV}}\right)^{-2.27} \left(\frac{g_{\ast}}{30}\right)^{\frac{3}{2}} \left(\frac{d_\alpha}{1.27}\right)^{-1}\left(\frac{\Omega_\textrm{DM}h^2}{ 0.12}\right)^{-1}~, \\
        \sin^2(2\theta)_\textrm{old}^{\rm ST2} &~=~ 1.15\times10^{-8} \left(\frac{m_s}{\textrm{keV}}\right)^{-2} \left(\frac{g_{\ast}}{30}\right)^{\frac{3}{2}}\left(\frac{d_\alpha}{1.27}\right)^{-1} \left(\frac{\Omega_\textrm{DM}h^2}{ 0.12}\right)^{-1}~,  \notag\\
        \sin^2(2\theta)_\textrm{old}^{\rm LRT} &~=~ 1\times10^{-3}\left(\frac{m_s}{\textrm{keV}}\right)^{-1}\left(\frac{d_\alpha}{1.13}\right)^{-1}\left(\frac{\Omega_\textrm{DM}h^2}{0.12}\right)^{-1}~.\notag
\end{align}

The way in which  $f_{\nu_s}$ approaches $f_{\nu_\alpha}$
with increasing mixing angle is quantified by the solution  to Eq.~\eqref{eq:Boltzmann} which we call ``non-linear" $f_{\nu_s-{\rm nl}}$~\cite{Rehagen:2014vna}, 
\begin{equation} 
\label{eq:nonlinearprod}
    f_{\nu_s-{\rm nl}}(\epsilon, T)= \left(1- e^{-K(\epsilon, T)}\right)  f_{\nu_\alpha} =  \left[ 1-\exp{ \left( -\dfrac{f_{\nu_s-{\rm lin}}(\epsilon, T)}{f_{\nu_\alpha}(\epsilon)}\right)} \right] {f_{\nu_\alpha}(\epsilon)}~.
\end{equation}
 Here, $K(\epsilon, T)$ is 
\begin{equation}
\label{K-definition}
    K(\epsilon, T) = \int_T^\infty  dT ~  \left(\dfrac{\Gamma_s (\epsilon, T)}{HT}\right)_\epsilon~= \dfrac{f_{\nu_s-{\rm lin}}(\epsilon, T)} {f_{\nu_\alpha}(\epsilon)}~,
\end{equation}
for all the cosmologies we consider except LRT, for which the upper limit of integration is $T_{\rm RH}$ (see below). Notice that the integral in Eq.~\eqref{K-definition} is performed while keeping $\epsilon$ constant. We assume that there are no sterile neutrinos present before non-resonant production takes place.
Eq.~\eqref{eq:nonlinearprod} can be easily verified to be the solution to Eq.~\eqref{eq:Boltzmann} by substitution. Our previous solution is readily recovered Eq.~\eqref{eq:nonlinearprod} as $f_{\nu_s-{\rm lin}}$ becomes much smaller than $f_{\nu_\alpha}$ and we then keep only the first non-trivial term in the exponential.

Eq.~\eqref{K-definition} corresponds to Eq.~(3.10) of Paper I, except that in Paper I we approximated the lower limit of integration with $T=0$. This is justified as the temperatures of interest in Eq.~\eqref{eq:nonlinearprod}, the lower limits of integration in Eq.~\eqref{K-definition},  are much lower than the temperature $T_{\rm max}$ at which the sterile neutrino production rate $(\partial f_{\nu_s}/\partial T)_{\epsilon}$ (neglecting $f_{\nu_s}$ in the right hand side of Eq.~\eqref{eq:Boltzmann}) has a sharp maximum. For the STD cosmology, $T_{\rm max}$ is
\begin{equation}
\label{eq:Stdmax}
    T_{\textrm{max}}^{\rm STD} = 145 \textrm{ MeV} \left(\frac{m_s}{\textrm{keV}}\right)^{\frac{1}{3}}\epsilon^{-\frac{1}{3}}~,
\end{equation}
and it is similar in the K, ST1 and ST2 cosmologies (see Eqs.~(3.8), (3.9), (A.2) and (A.3) of Paper I).
  This is a good approximation for the STD, ST1, ST2 and K cosmologies. In the LRT model, all of the sterile neutrino production is assumed to occur only during
the late standard cosmology phase, at $T< T_{\rm RH}$ below the reheating temperature. Thus, the upper limit of integration in Eq.~\eqref{K-definition} becomes $T_{\rm RH}$. As the maximum of the production happens very close to $T_{\rm RH}$, the lower limit of integration can again be taken to be $T=0$. This is the reason why the $f_{\nu_s-{\rm lin}}$ in Eq.~\eqref{eq:flin} are function of $\epsilon$ only (and not $T$). Therefore, from Eq.~\eqref{eq:nonlinearprod} we see that $f_{\nu_s-{\rm nl}}$ are also functions only of $\epsilon$.

Notice that $f_{\nu_s-{\rm nl}}$ in Eq.~\eqref{eq:nonlinearprod} approaches the active neutrino distribution as the linear solution $f_{\nu_s-{\rm lin}}$ grows larger than $f_{\nu_\alpha}$.~This $f_{\nu_s-{\rm lin}}$ is a non-physical solution of the Boltzmann equation due to not taking into account $f_{\nu_s}$ on the right hand side.

As we will now show, the function  $f_{\nu_s-{\rm nl}}$ departs significantly from the linear solution $f_{\nu_s-{\rm lin}}$ 
for mixing angles that are forbidden by the DM density condition  $\Omega_s < \Omega_{\rm DM}$ and by the upper limit $N_{\rm eff} < 3.4$ on the effective number of relativistic active neutrino species present during BBN. As these regions are already forbidden, the resulting limits of Paper I are unaffected by thermalization considerations.

In order to derive all limits that depend on the sterile neutrino number density $n_{\nu_s}$, one needs to integrate $f_{\nu_s-{\rm nl}}$ over momenta to obtain $n_{\nu_s}$. Following our previous notation we will denote the integration result as ``non-linear" number density $n_{\nu_s-{\rm nl}}$, and the number densities we presented before in Paper I as ``linear" $n_{\nu_s-{\rm lin}}$.  The integration needs to be performed numerically, unless the ratio $(f_{\nu_s-{\rm lin}}/f_{\nu_\alpha})$ is a constant independent of $\epsilon$. This is the case for the STD and ST2 cosmologies, where
\begin{equation} 
\label{eq:linearnumber}
    \left(\dfrac{n_{\nu_s-{\rm lin}}}{n_{\nu_\alpha}}\right)=
    \left(\dfrac{f_{\nu_s-{\rm lin}}}{f_{\nu_\alpha}}\right)
    \end{equation}
and we can obtain the exact solution for $n_{\nu_s-{\rm nl}}$,
\begin{equation}
\label{eq:nonlinearprod2}    
     \left(\frac{n_{\nu_s-{\rm nl}}}{n_{\nu_\alpha}}\right) = 1-e^{-(n_{\nu_s-{\rm lin}}/n_{\nu_\alpha})}~.
\end{equation}

Following our analysis of Paper I, we will proceed with an analytic treatment. We are going to find approximate analytic solutions for $n_{\nu_s-{\rm nl}}$ for the other cosmologies we consider in which Eq.~\eqref{eq:linearnumber}  does not hold, because the ratio $(f_{\nu_s-{\rm lin}}/f_{\nu_\alpha})$ depends on $\epsilon$. The exact solution would require integration over momentum of an exponential function of $\epsilon$. Since the dependence of the  $(f_{\nu_s-{\rm lin}}/f_{\nu_\alpha})$ ratio on $\epsilon$ is weak, our approximation is justified.

With the DM abundance $\Omega_{\rm DM}h^2 = 0.12$,  a fully thermalized sterile neutrino, with the relic number density of an active neutrino species, would constitute all of the DM if its mass is $m_s = 11.5\textrm{ eV}$. Thus, the DM limit $\Omega_s < \Omega_{\rm DM}$ does not restrict sterile neutrinos with $m_s < 11.5\textrm{ eV}$, since the number density of sterile neutrinos is at most equal to that of one active neutrino species. Above and close to $m_s = 11.5\textrm{ eV}$, taking into account the  non-linear solution $f_{\nu_s-{\rm nl}}$  modifies the DM density limit with respect to the results of Paper I.  

In order to find an approximate analytic solution for $n_{\nu_s-{\rm nl}}$, let us start by defining a pre-factor $C$ such that Eq.~(A.10) of Ref.~\cite{Gelmini:2019clw} for the sterile neutrino relic number density is 
\begin{equation}
\label{eq:Cdef}
n_{\nu_s-{\rm lin}} = C \sin^2 2\theta
\end{equation}
(i.e. $C$ includes all the factors independent of the active-sterile mixing angle). Then the non-linear solution $n_{\nu_s-nl}$ for the number density  satisfies 
\begin{equation} \label{eq:rhodef}
 \rho_{\rm DM}= \frac{m_s ~ n_{\nu_s-nl}}{\Omega_s/\Omega_{\rm DM}} = 11.5\textrm{ eV}~n_{\nu_\alpha}~.  
\end{equation}
We denoted the DM density limit obtained using $n_{\nu_s-{\rm lin}}$, as in Paper I, as ($\sin^2 2\theta)_{\rm old}$ that is a function of $m_s$ given in Eq.~\eqref{eq:thetalin}. Thus, we can now state Eq.~(A.25) of Paper I for the DM fraction in sterile neutrinos for the K and ST2 cosmologies (or specifically Eqs.~(A.26)  and (A.27) of Paper I) and  Eq.~(A.28) of Paper I for the same fraction for the LRT model, setting these fractions to 1, as
\begin{equation}
\label{eq:old-limit}
    1= \frac{n_{\nu_s-{\rm lin}} m_s}{\rho_{\rm DM} ~ (\Omega_s/\Omega_{\rm DM})}= \frac{C  (\sin^2 2\theta)_{\rm old} m_s}{\rho_{\rm DM} ~ (\Omega_s/\Omega_{\rm DM})}~.
\end{equation}
Using Eqs.~\eqref{eq:Cdef}, \eqref{eq:rhodef} and \eqref{eq:old-limit} we can relate  $n_{\nu_s-nl}$ with the Paper I DM limit ($\sin^2 2\theta)_{\rm old}$,
\begin{equation}
\label{eq:linearrelation}
    \frac{C(\sin^2 2\theta)_\textrm{old}}{n_{\nu_\alpha}} = \frac{({\Omega_s}/\Omega_{\rm DM})~\rho_{\rm DM}}{m_s n_{\nu_\alpha}} = \frac{(\Omega_s/\Omega_{\rm DM})~11.5\textrm{ eV}}{m_s}= \frac{n_{\nu_s-nl}}{n_{\nu_\alpha}}~.
\end{equation}
This allows to define $(\sin^2 2\theta)_\textrm{new}$ such that the ratio $(n_{\nu_s-{\rm nl}}/n_{\nu_\alpha})$ in Eq.~\eqref{eq:nonlinearprod2} satisfies Eq.~\eqref{eq:linearrelation} when  $(\sin^2 2\theta)_\textrm{new}$  is used in $n_{\nu_s-{\rm lin}}$ in the exponent in the same equation, so that $(n_{\nu_s-{\rm nl}}/n_{\nu_\alpha})= (\Omega_s/\Omega_{\rm DM})~ 11.5~ {\rm eV}/ m_s$. Hence, 
\begin{equation}
\label{eq:nonlinearprod3}
    \frac{(\Omega_s/\Omega_{\rm DM})~11.5\textrm{ eV}}{m_s} = 1 - \exp \left(-\frac{C(\sin^2 2\theta)_\textrm{new}}{n_{\nu_\alpha}}\right)~.
\end{equation}
Replacing  here $n_{\nu_\alpha}$  by $C(\sin^2 2\theta)_\textrm{old}m_s/(\Omega_s/\Omega_{\rm DM})~11.5\textrm{ eV}$ using  Eq.~\eqref{eq:linearrelation}, Eq.~\eqref{eq:nonlinearprod3} can be rearranged to give the new mixing angle for the DM density limit (plotted in the figures) in terms of the old mixing angle (see Eqs.~(A.29) to (A.32) of Paper I)
\begin{equation}
\label{eq:newoverdenselimit}
    (\sin^2 2\theta)_\textrm{new} = (\sin^2 2\theta)_\textrm{old}~ \dfrac{m_s}{(\Omega_s/\Omega_{\rm DM})~11.5 ~{\rm eV}}~ \ln\left[\dfrac{m_s}{m_s - (\Omega_s/\Omega_{\rm DM})~ 11.5\textrm{~eV}}\right]~.
\end{equation}
Taking $(\Omega_s/\Omega_{\rm DM})=1$ this is the boundary of the dark gray regions where $\Omega_s > \Omega _{\rm DM}$ shown in  Fig.~\ref{fig:lrDWlim} and Fig.~\ref{fig:allDWlim}. Except in a region close to or below $m_s = 11.5$ eV, which is rejected by the (cyan) $N_{\rm eff}$ BBN limit, the present DM density limits are the same as those in Paper I. Thus the allowed regions have not changed.

\section{Thermalization}

 The production of sterile neutrinos saturates when they thermalize,  when $f_{\nu_s}=f_{\nu_\alpha}$, and thus the right hand side of the Boltzmann equation Eq.~\eqref{eq:Boltzmann} is  equal to zero. In
 Fig.~\ref{fig:allDWlim}, the region of thermalization where $\Gamma/H|_{T_\textrm{max}} \geq 1$ is demarcated  with a solid blue line at its lower boundary. When the maximum production rate $\Gamma(T_{\rm max})$ stays roughly equal to or larger than the Hubble parameter for a significant period of time, a substantial amount of sterile neutrinos are produced and the population is nearly or fully thermalized. 
 
\begin{figure}[htb]
\begin{center}
\includegraphics[trim={5mm 0mm 40 0},clip,width=.475\textwidth]{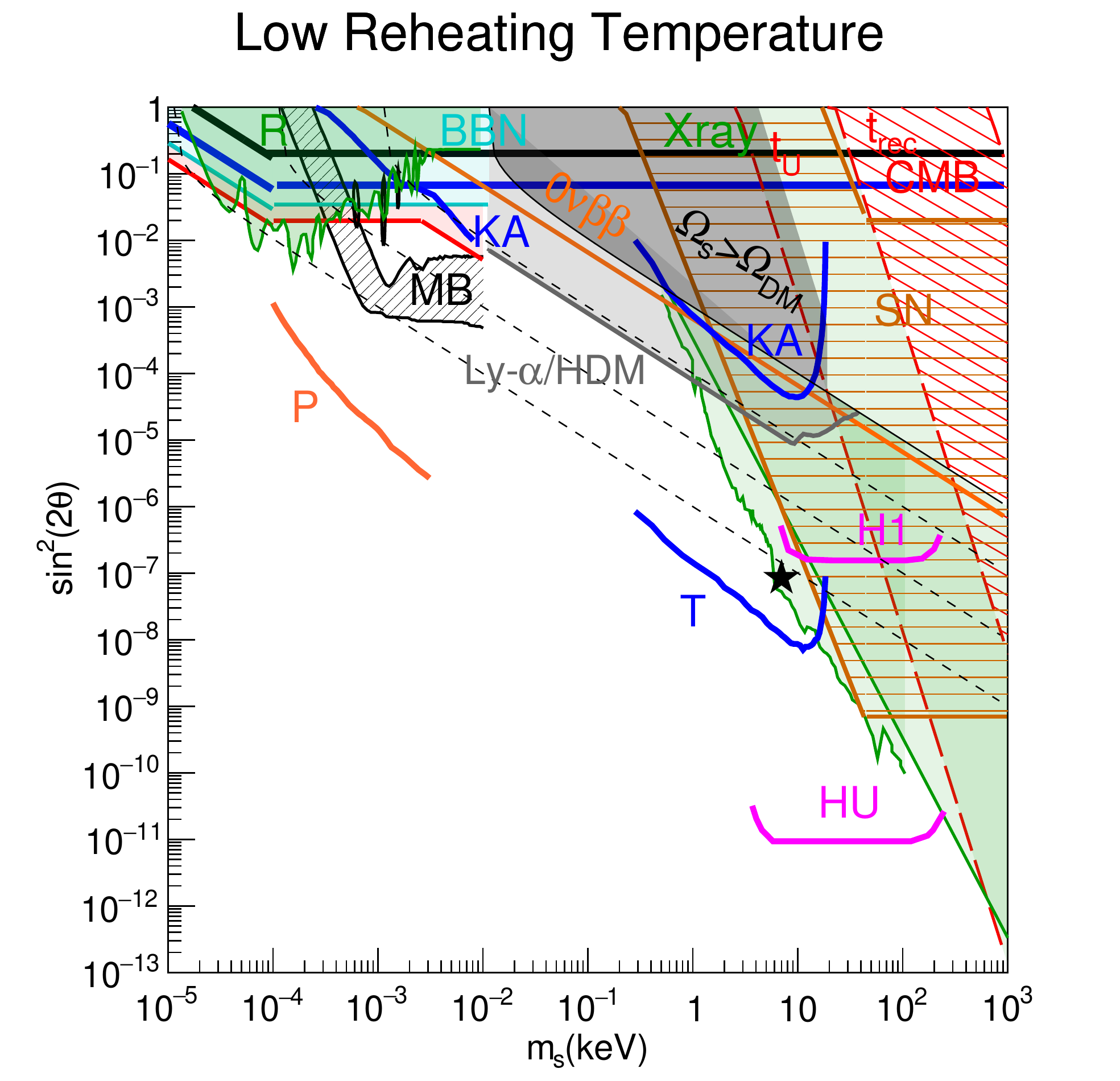}
\caption{ \label{fig:lrDWlim} Present relic abundance, limits and regions of interest in the mass-mixing space of a  $\nu_s$ mixed with $\nu_e$, for the LRT cosmology with $T_{\rm RH} = 5$ MeV~\cite{Gelmini:2004ah}. $g_\ast =10.75$ is used for $m_s<11.5\textrm{ eV}$, and $g_\ast = 30$ above. Shown are lines of $\Omega_s/\Omega_{\rm DM}=$ 1 (black solid line), $10^{-1}$, $10^{-2}$ and $10^{-3}$ (black dotted lines), the forbidden region $\Omega_s/\Omega_{\rm DM} > 1$  (shaded in dark gray), lifetimes $\tau= t_U$, $t_{\rm rec}$ and $t_{\rm th}$ of Majorana $\nu_s$ (long dashed red lines), the region (SN) disfavored by supernovae~\cite{Kainulainen:1990bn} (horizontally hatched in brown), the location  of the 3.5 keV X-ray signal~\cite{Bulbul:2014sua,Boyarsky:2014jta} (black star). The regions rejected by reactor neutrino (R) experiments (Daya Bay~\cite{An:2016luf}, Bugey-3~\cite{Declais:1994su} and  PROSPECT~\cite{Ashenfelter:2018iov}) shown in green, limits on $N_{\rm eff}$ during BBN~\cite{Tanabashi:2018oca} (BBN) in cyan,  Lyman-alpha limits~\cite{Baur:2017stq} (Ly-$\alpha$/HDM) shaded in light gray, X-ray limits~\cite{Ng:2019gch,Perez:2016tcq,Neronov:2016wdd} including DEBRA~\cite{Boyarsky:2005us}  (Xray) in green, $0\nu\beta\beta$ decays~\cite{KamLAND-Zen:2016pfg}  ($0\nu\beta\beta$) in orange  and CMB spectrum distortions~\cite{Fixsen:1996nj} (CMB) diagonally hatched in red. Current/future sensitivity of KATRIN (KA) in the keV~\cite{Mertens:2018vuu} and  eV~\cite{megas:thesis}  mass range, its TRISTAN upgrade in 3 yr (T)~\cite{Mertens:2018vuu} shown by blue solid lines. Magenta solid lines  show the reach of the  phase 1A (H1) of HUNTER, and its upgrade (HU)~\cite{Smith:2016vku}.  The 4-$\sigma$ band of compatibility with LSND and MiniBooNE results (MB) in Fig.~4 of~\cite{Aguilar-Arevalo:2018gpe} is shown densely hatched in black.  The three black vertical elliptical  contours are the regions allowed at 3-$\sigma$ by DANSS~\cite{Alekseev:2018efk} and NEOS~\cite{Ko:2016owz} data in Fig.~4 of~\cite{Gariazzo:2018mwd}). Orange solid lines show the reach of PTOLEMY  for 100 g-yr (P) exposure (from Figs. 6 and 7 of~\cite{Betti:2019ouf}). See Paper I for details. The thick blue line represents the thermalization condition $f_{\nu_s-\textrm{lin}}=f_{\nu_\alpha}$ and the thick black line shows $f_{\nu_s-\textrm{lin}}=3f_{\nu_\alpha}$. Notice that the LRT model goes into the standard cosmology, thus all limits become those standard, when $T_{\rm max} < T_{\rm RH}$ = 5 MeV, i.e. for $m_s < 0.1$ eV. The red lines in the upper-left hand corner denotes the Planck 2018 $\Delta N_\textrm{eff}$ and $m_\textrm{eff}$ bounds~\cite{Aghanim:2018eyx}.
}
\end{center}
\end{figure}

\begin{figure}[htb]
\begin{center}
\includegraphics[trim={5mm 0mm 40 0},clip,width=.475\textwidth]{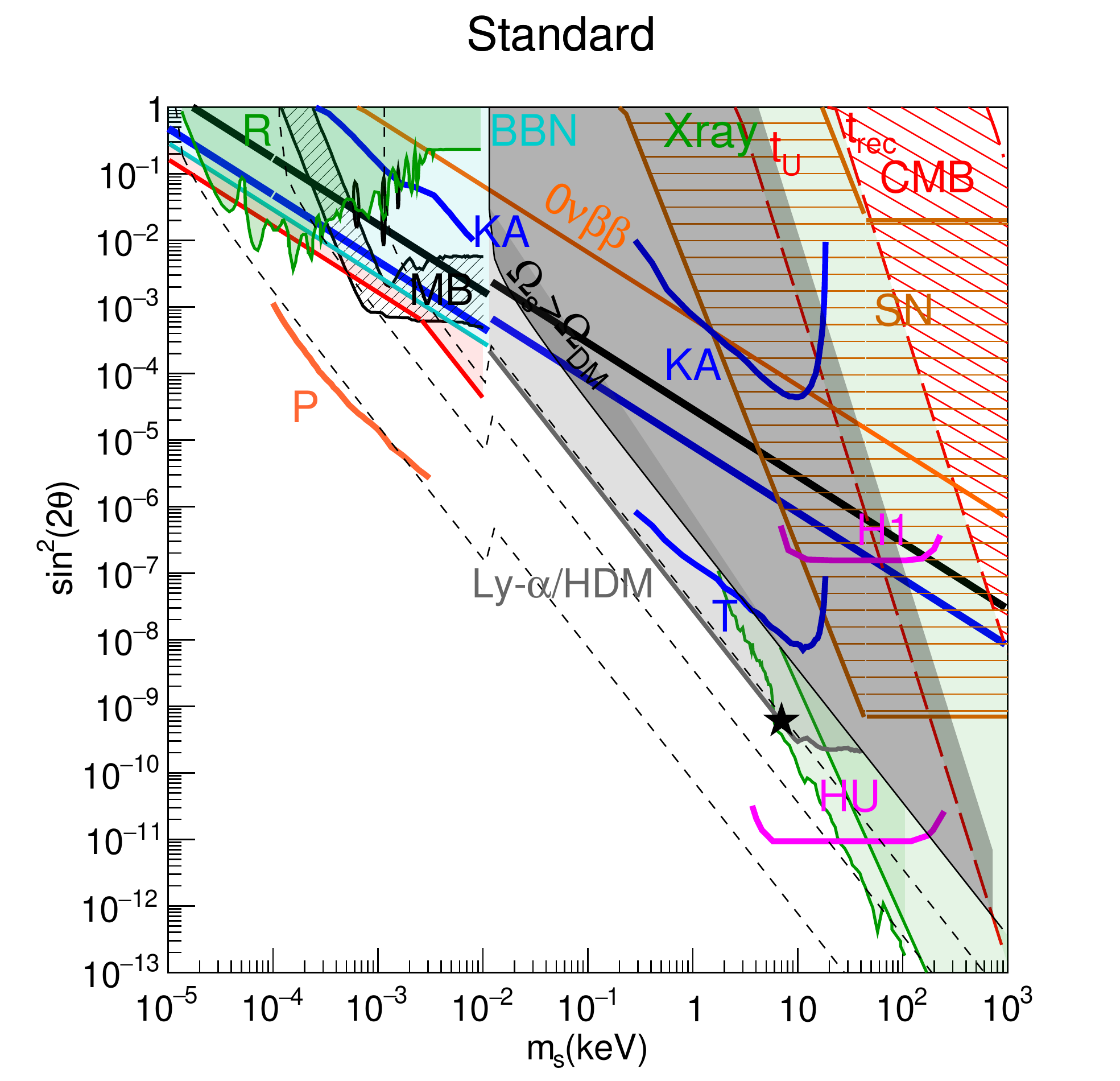}
\includegraphics[trim={5mm 0mm 40 0},clip,width=.475\textwidth]{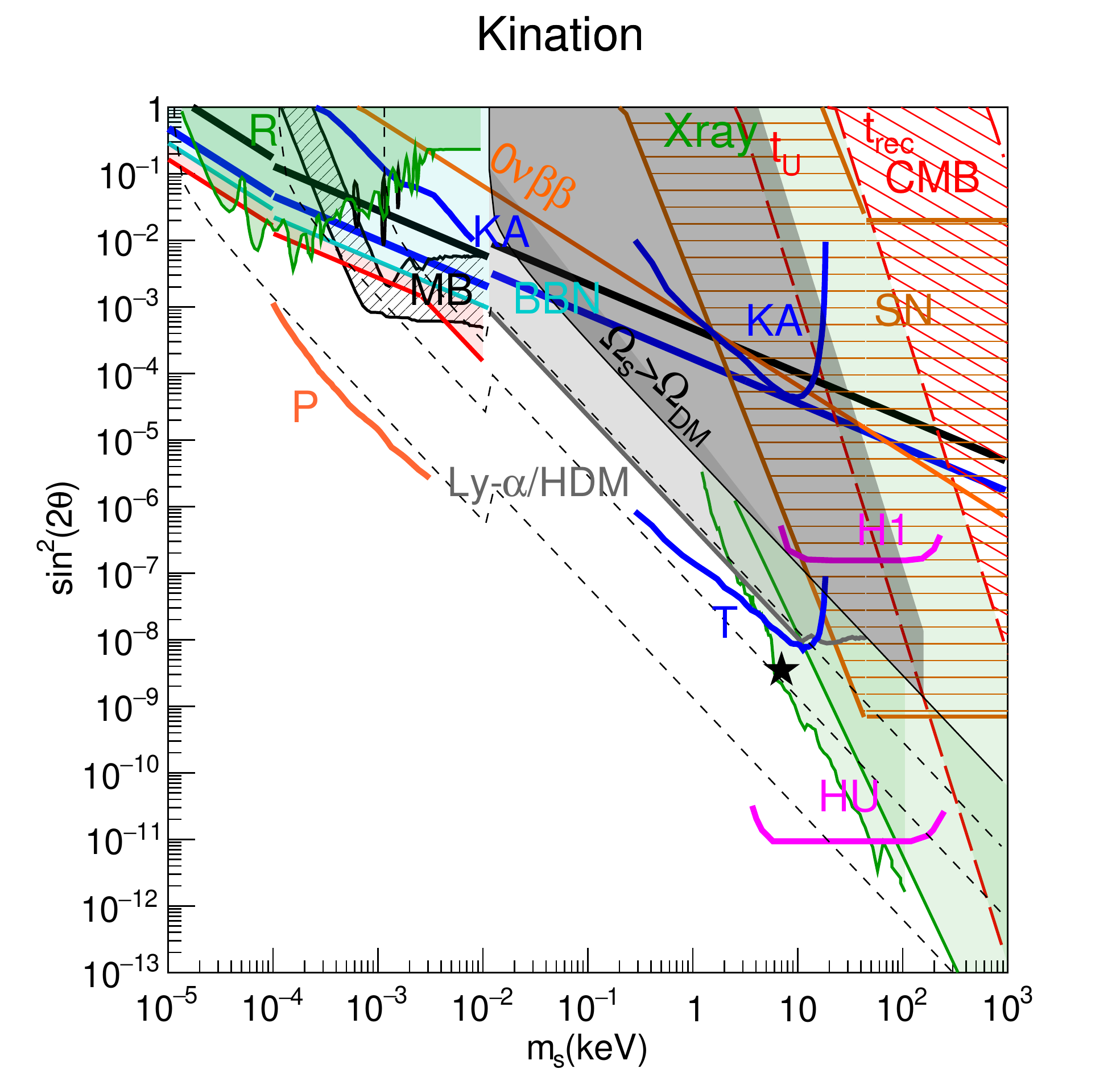}
\includegraphics[trim={5mm 0mm 40 0},clip,width=.475\textwidth]{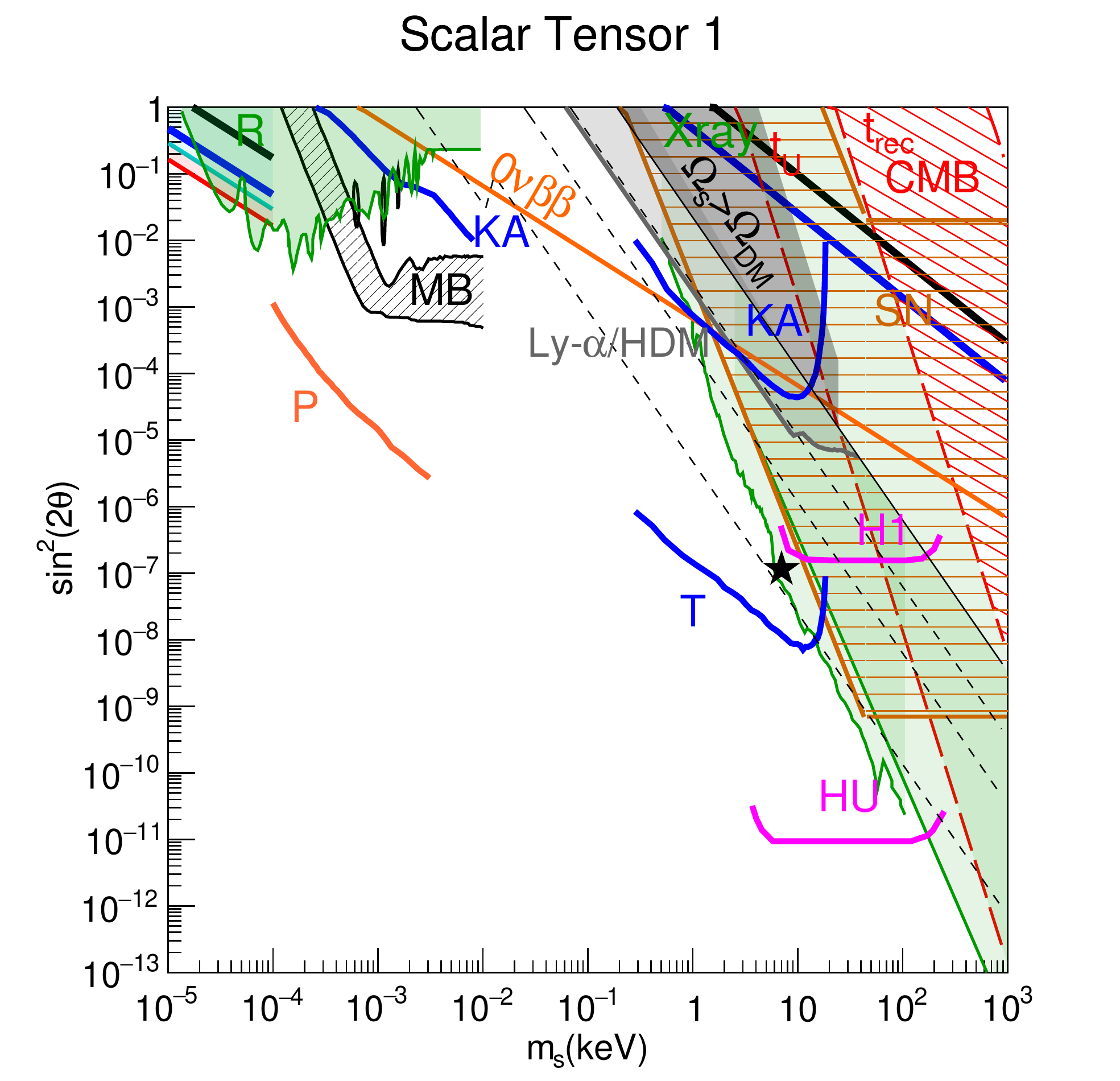}
\includegraphics[trim={5mm 0mm 40 0},clip,width=.475\textwidth]{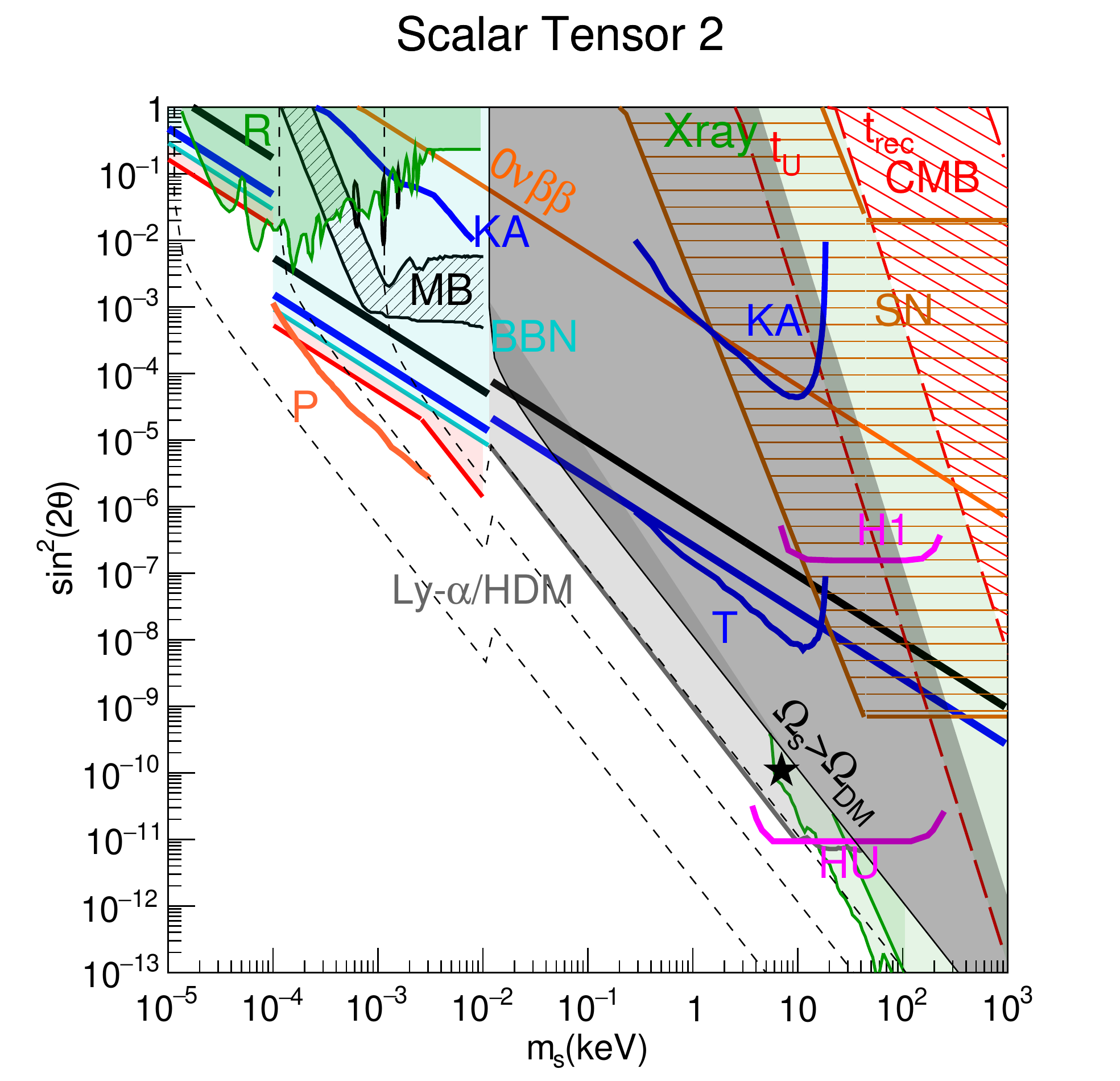}
\caption{  \label{fig:allDWlim}  Present relic abundance, limits and regions of interest for standard, kination and scalar-tensor cosmologies. See Fig.~\ref{fig:lrDWlim} caption. Here the thick blue line represents the condition $\Gamma/H|_{T_\textrm{max}} = 1$ that coincides with $f_{\nu_s}=f_{\nu_\alpha}$ as in Fig.~\ref{fig:lrDWlim}. The discontinuity in the limits at $m_s=11.5$ eV is due to our use of just two values for $g_\ast$, 10.75 below and 30 above (see explanations in the text). The red lines in the upper left hand corner show the Planck 2018 $\Delta N_\textrm{eff}$ and $m_\textrm{eff}$ bounds. Notice that all cosmologies go into the standard cosmology, thus all limits become those standard, when $T_{\rm max} < T_{\rm tr}$ = 5 MeV, i.e. for $m_s < 0.1$ eV.
}
\end{center}
\end{figure}

 To compute the production rates and momentum distributions we use as the characteristic momentum $\epsilon = \langle \epsilon\rangle$.  $\langle \epsilon\rangle$ is the average value of E/T for each cosmology (see Eqs.~(3.27) and (3.28) of Paper I)
 \begin{equation}
\label{eq:nonreseps}
\langle\epsilon\rangle
= \left \{
  \begin{tabular}{cl}
  3.15, & ~~~\text{STD}\\
  3.47, & ~~~\text{K} \\
  2.89, & ~~~\text{ST1} \\
  3.15, & ~~~\text{ST2}\\
  4.11, & ~~~\text{LRT}\\
  \end{tabular}
\right .\
\end{equation}
In contrast to Paper I, except for LRT we use two values of the effective number of degrees of freedom contributing to the radiation density in $H$, $g_\ast = 10.75$ for $m_s < 11.5\textrm{ eV}$ and $g_\ast = 30$ for $m_s>11.5\textrm{ eV}$. This choice allows to better approximate  the evolution of $g_\ast$ with temperature~\cite{Husdal:2016haj,Borsanyi:2016ksw,Drees:2015exa}. We have chosen $m_s=11.5$ eV as the mass where $g_\ast$ changes, because for this mass $T_\textrm{max}\simeq 20$ MeV and this is the temperature above which $g_\ast$ starts increasing from its value of 10.75.   In Paper I we had adopted for simplicity $g_\ast=30$ throughout the entire mass range, except for the $N_ {\rm eff}$  BBN limit,  which is particularly relevant for light sterile neutrinos, and the LRT cosmology, for which we used 10.75. Here we instead adopt $g_\ast = 10.75$ for all our calculations with $m_s < 11.5$ eV as this value is more appropriate to the sterile neutrino production and thermalization at the eV scale, specifically in the regions where possible LSND, MiniBooNE, DANSS and NEOS sterile neutrino detection signals have been suggested. Our choice of using two distinct values of $g_\ast$ results in an artificial discontinuity\footnote{Had we instead considered the true value of $g_\ast$ that is a continuous function of temperature, such discontinuity would be absent.} at $m_s = 11.5$ eV in all the limits in  Fig.~\ref{fig:allDWlim}. In the LRT cosmology, production  happens only at $T<$  5 MeV, for which $g_\ast = 10.75$, for all sterile neutrinos. Thus there are no discontinuities at $m_s = 11.5$ eV in the BBN $N_ {\rm eff}$ and thermalization (cyan, blue and black) limits in Fig.~\ref{fig:lrDWlim}.

Notice that all cosmologies go into the standard cosmology, thus all limits become those standard, when $T_{\rm max} < T_{\rm tr}$ = 5 MeV, i.e. for $m_s < 0.1$ eV. Given our approximations of considering a sharp transition of all cosmologies into the standard one at $T_{\rm tr}$, and assuming the sterile neutrino production happens at $T_{\rm max}$,  this 
results in a discontinuity at $m_s\simeq 0.1\textrm{ eV}$ in the limits in in Fig.~\ref{fig:lrDWlim} and~\ref{fig:allDWlim}, which had not been included in Paper I (as it affects a very small portion of the whole mass range we considered). In a more careful treatment, the limits would smoothly transition from the non-standard to the standard ones.

Solving for $\sin^2 2\theta$ from  the condition $\Gamma/H|_{T_\textrm{max}} = 1$ we obtain the following thermalization limits (the solid thick blue lines in Fig.~\ref{fig:allDWlim}), 
\begin{align}
& \textrm{for STD:}&  &(\sin^2 2\theta)_{\rm th} = 4.86\times10^{-3}\left(\frac{d_\alpha}{1.27}\right)^{-1}\left(\frac{m_s}{\textrm{eV}}\right)^{-1}\left(\frac{g_\ast}{10.75}\right)^{\frac{1}{2}}~, \\
& \textrm{for K:}&  &(\sin^2 2\theta)_{\rm th} = 1.52\times10^{-2}\epsilon^{-\frac{1}{3}}\left(\frac{d_\alpha}{1.27}\right)^{-1}\left(\frac{m_s}{\textrm{eV}}\right)^{-\frac{2}{3}}\left(\frac{g_\ast}{10.75}\right)^{\frac{1}{2}}~, \\
& \textrm{for ST1:}&  &(\sin^2 2\theta)_{\rm th} = 1.38\times10^{3} \epsilon^{0.27}\left(\frac{d_\alpha}{1.27}\right)^{-1}\left(\frac{m_s}{\textrm{eV}}\right)^{-1.27}\left(\frac{g_\ast}{10.75}\right)^{\frac{1}{2}}~, \\
& \textrm{and for ST2:}&  &(\sin^2 2\theta)_{\rm th} = 1.56\times10^{-4}\left(\frac{d_\alpha}{1.27}\right)^{-1}\left(\frac{m_s}{\textrm{eV}}\right)^{-1}\left(\frac{g_\ast}{10.75}\right)^{\frac{1}{2}}~. 
\end{align} 
We have confirmed that these limits (derived from $\Gamma/H|_{T_\textrm{max}} = 1$) practically coincide with those corresponding to $f_{\nu_s-\textrm{lin}}= f_{\nu_\alpha}$ for the mentioned cosmologies, which we thus do not display separately in Fig.~\ref{fig:allDWlim}.

For the LRT model, considering that the maximum production rate is at $T_{\rm RH}=$ 5 MeV, we could be tempted to use $\Gamma/H|_{T_\textrm{RH}} = 1$
 as the condition for thermalization. However, when employing this condition throughout the whole range of integration in $T$, from 0 to $T_{\rm RH}$, to obtain $K(\epsilon, T)$, the integrand is smaller than 1. Thus, this is not a good condition of thermalization for this model. Since  the thermalization condition based on $\Gamma/H$  coincides with the condition $f_{\nu_s-\textrm{lin}}= f_{\nu_\alpha}$ in all the other models we consider, we thus adopt $f_{\nu_s-\textrm{lin}}= f_{\nu_\alpha}$ as the condition for thermalization in the LRT model. This condition translates into  
\begin{equation}
\label{eq:lrtgammah}
    (\sin^2 2\theta)_{\rm th} = 2.78 \times 10^{-1} ~\epsilon^{-1}~,
\end{equation}
which is shown with the thick blue line in Fig.~\ref{fig:lrDWlim}.

Notice that we have considered the condition $\Gamma/H > 1$ for chemical equilibrium of sterile neutrinos, since the rate $\Gamma$ we used is the production rate.  Kinetic equilibrium happens at larger mixing angles than  chemical equilibrium. The reason for this is that the sterile neutrino scattering rate contains an extra $\sin^2 \theta$ factor over the production rate. Thus, sterile neutrinos that are not in chemical equilibrium (i.e. for which the production rate is $\Gamma< H$) are also not in kinetic equilibrium, they are decoupled from the thermal bath.

On the thick blue lines in the figures, $f_{\nu_s-\textrm{nl}} = (1-e^{-1})f_{\nu_\alpha} = 0.63 f_{\nu_\alpha}$. In Fig.~\ref{fig:lrDWlim} and Fig.~\ref{fig:allDWlim} we also display with a solid black line where $f_{\nu_s-\textrm{lin}}=3 f_{\nu_\alpha}$, and thus $f_{\nu_s-\textrm{nl}} = (1-e^{-3})f_{\nu_\alpha} = 0.95 f_{\nu_\alpha}$, where nearly full thermalization occurs. Above this black line, the sterile neutrino momentum distribution rapidly becomes $f_{\nu_s}=f_{\nu_\alpha}$ with increased mixing (i.e.~the right hand side of the Boltzmann equation Eq.~\eqref{eq:Boltzmann} goes to zero). The equations of the thick black line in the figures are:
\begin{align}
& \textrm{for STD:}&  &\sin^2 2\theta = 1.73\times10^{-2}\left(\frac{d_\alpha}{1.27}\right)^{-1}\left(\frac{m_s}{\textrm{eV}}\right)^{-1}\left(\frac{g_\ast}{10.75}\right)^{\frac{1}{2}}~, \\
& \textrm{for K:}&  &\sin^2 2\theta = 4.29\times10^{-2}\epsilon^{-\frac{1}{3}}\left(\frac{d_\alpha}{1.27}\right)^{-1}\left(\frac{m_s}{\textrm{eV}}\right)^{-\frac{2}{3}}\left(\frac{g_\ast}{10.75}\right)^{\frac{1}{2}}~, \\
& \textrm{for ST1:}&  &\sin^2 2\theta = 5.27 \times 10^3 \epsilon^{0.27}\left(\frac{d_\alpha}{1.27}\right)^{-1}\left(\frac{m_s}{\textrm{eV}}\right)^{-1.27}\left(\frac{g_\ast}{10.75}\right)^{\frac{1}{2}}~,
\\
& \textrm{for ST2:}&  &\sin^2 2\theta = 5.52\times10^{-4}\left(\frac{d_\alpha}{1.27}\right)^{-1}\left(\frac{m_s}{\textrm{eV}}\right)^{-1}\left(\frac{g_\ast}{10.75}\right)^{\frac{1}{2}}~, \\
& \textrm{and for LRT:}&  &\sin^2 2\theta = 8.34\times10^{-1} \epsilon^{-1}~. 
\end{align} 
Eqs.~\eqref{eq:linearnumber} and~\eqref{eq:nonlinearprod2} emply that  $f_{\nu_s-{\rm lin}}=3 f_{\nu_\alpha}$ corresponds to $n_{\nu_s-{\rm nl}}/n_{\nu_\alpha} \simeq 0.95$, which leads to $\rho_s = \rho_{\rm DM}$  for $m_s= 11.5$ eV. In fact, in the figures the thick blue line intersects the DM density limit near $m_s=$ 11.5 eV, as expected.

\section{Bounds}
We include here the same bounds detailed in Paper I with a few modifications. As we explain below, due to solving for the non-linear number densities as described in the preceding sections, both the BBN $\Delta N_\textrm{eff}$ and Ly-$\alpha$ bounds move here to larger mixings with respect to those in Paper I, and also add here the CMB $\Delta N_\textrm{eff}$ and $m_\textrm{eff}$ bounds~\cite{Aghanim:2018eyx}, which we had neglected  in Paper I because they are very close to the BBN $\Delta N_\textrm{eff}$ limit).

To derive the Lyman-$\alpha$ bound we used the $2\textrm{-}\sigma$ warm DM limit from SDSS+XQ+HR in Fig. 6 of Ref.~\cite{Baur:2017stq}, which has an asymptote of $(\Omega_s/\Omega_{\rm DM})\lesssim 0.08$ for small sterile neutrino masses. This limit is given in terms of $m_\textrm{therm}$ which can be converted to limits on $m_s$ using~\cite{Viel:2005qj} $m_s = 4.46~{\rm keV}\left(\langle\epsilon\rangle/3.15\right)
\left(m_\textrm{therm}/\textrm{keV}\right)^{\frac{4}{3}}
\left(T_{\nu_s}/T_{\nu_\alpha}\right)\left(0.12/(\Omega_{s}~ h^2)\right)^{\frac{1}{3}}$. We apply Eq.~\eqref{eq:newoverdenselimit}  with $(\Omega_s/\Omega_{\rm DM}) = 0.08$ replacing $(\sin^2 2\theta)_\textrm{old}$ by the Lyman-$\alpha$ limits in Paper I, to obtain the present Lyman-$\alpha$ bounds. The Lyman-$\alpha$ limits are shown up to their intersection with THE BBN $N_{eff}$ bounds. Using Eq.~\eqref{eq:nonlinearprod2} we obtain the BBN $\Delta N_\textrm{eff} \leq 0.4$~\cite{Tanabashi:2018oca} limit, which translates into $n_{\nu_s-{\rm lin}}/n_{\nu_a} \leq 0.51$.

We apply  for $m_s \lesssim 10\textrm{ eV}$ the combined CMB $\Delta N_\textrm{eff}$ and $m_\textrm{eff}$~\cite{Aghanim:2018eyx}, for sterile neutrinos which are respectively relativistic and becoming non-relativistic close recombination. The current 95\% Planck 2018 limits in Eq.~(70a) of Ref.~\cite{Aghanim:2018eyx} are\footnote{While these bounds were formulated for thermally produced sterile neutrinos, they are expected to be reasonably accurate for other models~\cite{2016A&A...594A..13P}. We thus apply them to all cosmologies.}
\begin{equation}
\label{eq:cmblimits}
    N_\textrm{eff}<3.29,~~~~~~~ m_\textrm{eff}<0.65\textrm{ eV}~.
\end{equation}
Using the definitions 
$N_\textrm{eff}= 3.04 + (\rho_{\nu_s}/\rho_{\nu_\alpha})$ and $m_\textrm{eff} = n_{\nu_s} m_s/n_{\nu_a}$~\cite{Rehagen:2014vna,Aghanim:2018eyx}
with $\rho_{\nu_s}/\rho_{\nu_\alpha} = (\langle\epsilon\rangle ~  n_{\nu_s-{\rm nl}})/ (3.15~ n_{\nu_a})$ and $n_{\nu_s}=n_{\nu_s-{\rm nl}}$, and replacing in Eq~\eqref{eq:nonlinearprod2} the upper limits on $n_{\nu_s-{\rm nl}}$ derived from the $N_\textrm{eff}$ and the $m_\textrm{eff}$ limits  we get respectively
\begin{equation}
\label{eq:cmblimits2}
    \frac{n_{\nu_s-{\rm lin}}}{n_{\nu_\alpha}}< \ln\left(\frac{1}{1-0.25 (3.15/\langle\epsilon\rangle)}\right)\simeq 0.29,~~~~~~~~\frac{n_{\nu_s-{\rm lin}}}{n_{\nu_a}}<-\ln\left[1-\frac{0.65\textrm{ eV}}{m_s}\right]~.
\end{equation}
Using now Eqs. (3.18), (3.20), (A.12), (A.14), and (A.16) of Paper I for $n_{\nu_s-{\rm lin}}$, we obtain the upper limits on the mixing angle
shown with red solid lines in the upper left hand corners of Figs.~\ref{fig:lrDWlim} and~\ref{fig:allDWlim}  for $m_s<10\textrm{ eV}$. The $m_\textrm{eff}$ bound becomes more restrictive than the $\Delta N_\textrm{eff}$ bound for $m_s>3\textrm{ eV}$, which causes the change in slope of the red lines. As it is clear from the figures, these CMB limits are very close to the BBN $N_\textrm{eff}$ (cyan) limits, thus do not change significantly the allowed parameter regions (as we argued  in Paper I to neglect them).

\section{Concluding Remarks}

We have considered the approach of sterile neutrinos to thermalization that happens for large enough active-sterile mixing angles. We showed that the allowed regions of parameter space found in Paper I are not affected by these considerations.  In particular, the interesting region in which there are several suggested potential signals of a light sterile neutrino with mass close to 1 eV are free from cosmological bounds in the ST1 and LRT cosmologies. 

\acknowledgments
\addcontentsline{toc}{section}{Acknowledgments}
 
The work of G.B.G., P.L. and V.T. was supported in part by the U.S. Department of Energy (DOE) Grant No. DE-SC0009937.

\bibliography{sternumodcos}
\bibliographystyle{JHEP}

\end{document}